\documentclass{article}
\usepackage{amsfonts,amssymb, amsmath,mathrsfs}

\DeclareMathAlphabet{\mathpzc}{OT1}{pzc}{m}{it}

\def\nn{\nonumber }
\def\bq{ \begin{equation} }
\def\eq{ \end{equation} }
\def\ben{ \begin{eqnarray} }
\def\en{ \end{eqnarray} }

\def\e{{\rm e}}
\def\on#1#2{\mathop{\vbox{\ialign{##\crcr\noalign{\kern2pt}
$\scriptstyle{#2}$\crcr\noalign{\kern2pt\nointerlineskip}
\kern-2pt$\hfil\displaystyle{#1}\hfil$\crcr}}}\limits}

\newtheorem{prop}{Proposition}

\newtheorem{remark}{Remark}
\newenvironment{rem}{\begin{remark} \rm}{\end{remark}}
\begin{document}

\title{The Poisson bracket compatible
with the classical reflection equation algebra}
\author{
A. V. Tsiganov\\
St.Petersburg State University, St.Petersburg, Russia\\
\it\small e--mail: tsiganov@mph.phys.spbu.ru}

 \date{}
\maketitle

{\small
We introduce a family of compatible Poisson brackets
on the space of $2\times 2$ polynomial matrices, which contains the reflection equation algebra bracket. Then we use it to derive a multi-Hamiltonian structure for a set of
integrable systems that includes the $XXX$ Heisenberg magnet with boundary conditions, the generalized Toda lattices and the Kowalevski top.}

\section{Introduction}
\setcounter{equation}{0}
In this paper we study a class of finite-dimensional Liouville integrable systems described
by the representations of the quadratic $r$-matrix Poisson algebra:
\ben
\{\,\on{T}{1}(\lambda),\,\on{T}{2}(\mu)\}_0&=& [r(\lambda-\mu),\,
\on{T}{1}(\lambda)\on{T}{2}(\mu)\,]
\label{rrpoi}\\
&+&
\on{T}{1}(\lambda)r(\lambda+\mu)\on{T}{2}(\mu)-\on{T}{2}(\mu)r(\lambda+\mu)
\on{T}{1}(\lambda)
\,, \nn
\en
where $\on{T}{1}(\lambda)=T(\lambda)\otimes \mathrm I\,,~\on{T}{2}(\mu)=\mathrm I\otimes T(\mu)$ and $r(\lambda,\mu)$ is a classical $r$-matrix.

The reflection equation algebra (\ref{rrpoi}) appeared in the quantum inverse scattering method \cite{skl88}. Its representations play an important role in the classification and studies of classical integrable systems (see, for instance, \cite{ts04,kuzts89a,kuzts89,skl88} and references therein).

 The main result of this paper  is construction of the Poisson brackets $\{.,.\}_1$ compatible with the bracket $\{.,.\}_0$ (\ref{rrpoi})  in the simplest case of the $4\times4$ rational $r$-matrix
\bq
r(\lambda-\mu)=\dfrac{-\eta}{\lambda-\mu}\,\Pi,\qquad \Pi=\left(\begin{array}{cccc}
 1 & 0 & 0 & 0 \\
 0 & 0 & 1 & 0 \\
 0 & 1 & 0 & 0 \\
 0 & 0 & 0 & 1
\end{array}\right)\,,\qquad \eta\in\mathbb C,\label{rr}
\eq
and $2\times2$ matrix $T(\lambda)$, which depends polynomially on
the parameter $\lambda$
\bq\label{22T}
 T(\lambda)=\left(\begin{array}{cc}
 A (\lambda)& B (\lambda)\\
 C(\lambda) & A(-\lambda)
\end{array}\right), \qquad \mbox{\rm deg} T(\lambda)=\left(\begin{array}{cc}
 2n+1& 2n+1\\
 2n-1 & 2n+1
\end{array}\right).
\eq
Coefficients of the entries
\bq
\begin{array}{l}
A(\lambda)=\alpha\,\lambda^{2n+1} +A_{2n}\,\lambda^{2n}+A_{2n-1}\,\lambda^{2n-1}\ldots+A_0,\\
\\
B(\lambda)=\lambda^{2n+1}
+ B_{n}\lambda^{2n-1}+ B_{n-1}\lambda^{2n-3} \ldots+ B_1\lambda,\\
\\
C(\lambda)=C_{n}\lambda^{2n-1}+\ldots+C_{2}\lambda^{3}+C_1\lambda,
\end{array}
\label{asymp1}
\eq
are generators of the quadratic Poisson algebra (\ref{rrpoi}).
The leading coefficient $\alpha$ and   $2n+1$ coefficients of the $\det T(\lambda)$
\bq
d(\lambda)=\mathrm{det}\,T(\lambda)=Q_{2n}\lambda^{4n}+Q_{2n-1}\lambda^{4n-2}+\cdots+Q_{0}\,.
\label{Acentre}
\eq
are Casimirs of the bracket (\ref{rrpoi}). Therefore, we have a $4n+1$-dimensional space of the coefficients
\bq
A_0,\ldots, A_{2n},~B_1,\ldots, \ B_{n},~ \
C_1,\ldots, \ C_{n} \label{var1}
\eq
with $2n+1$ Casimir operators $Q_i$, leaving us with $n$ degrees of freedom.

The use of the algebra (\ref{rrpoi}) for the theory of the integrable systems is based on the following construction of commutative subalgebras \cite{skl88,sokts}.
Let us introduce boundary matrix
\bq
 K(\lambda)=\left(\begin{array}{cc}
 \mathcal A(\lambda) & 0 \\
 \mathcal C(\lambda) & \mathcal D(\lambda)
\end{array}\right)\,\label{KK}
\eq
where entries $\mathcal A(\lambda),\mathcal D(\lambda)$ are polynomials with numerical coefficients and $\mathcal C(\lambda)$ is arbitrary polynomial  on $\lambda$.
If the polynomial
\[\tau(\lambda)=\mbox{\rm tr}K(\lambda) T(\lambda)=\sum \tau_k\lambda^k\]
has $n$ independent dynamical coefficients $H_1,\ldots,H_n$ only, then
\[
\{\tau(\lambda),\tau(\mu)\}_0=0,\qquad\Rightarrow\qquad \{H_i,H_j\}_0=0,\qquad i,j=1,\ldots,n.
\]
These Poisson involutive integrals of motion $H_i$ define the Liouville integrable systems, which are our generic models for the whole paper.

\section{The compatible bracket}
\setcounter{equation}{0}

The Poisson brackets $\{.,.\}_{0}$ and $\{.,.\}_{1}$ are compatible if every linear combination of them is still a Poisson bracket \cite{fp02,mag97}.
\begin{prop}
The bracket (\ref{rrpoi}) belongs to the following family of
compatible Poisson brackets:
{\setlength\arraycolsep{1pt}
\ben
\{B(\lambda),B(\mu)\}_k&=&0,\quad k=0,1,\nn\\
\nn\\
\{A(\lambda),A(\mu)\}_k&=&
\frac{\scriptstyle \eta}{\scriptstyle \lambda+\mu}
\bigr(\mu^{2k}B(\lambda)C(\mu)-\lambda^{2k}B(\mu)C(\lambda)\bigl),\nn\\
\nn\\
\{C(\lambda),C(\mu)\}_k&=&
2\eta\Bigl(\rho_k(-\mu)A(\mu)-\rho_k(\mu)A(-\mu)\Bigr)C(\lambda)\nn\\
&-&2\eta\Bigl(\rho_k(-\lambda)A(\lambda)-\rho_k(\lambda)A(-\lambda)\Bigr)C(\mu)
,
\nn\\
\nn\\
\{A(\lambda),B(\mu)\}_k&=&
\frac{\scriptstyle \eta}{\scriptstyle\lambda-\mu}\Bigl( \lambda^{2k}A(\lambda)B(\mu)-\mu^{2k}A(\mu)B(\lambda)\Bigr)\nn\\
&+&\frac{\scriptstyle\eta}{\scriptstyle\lambda+\mu}\Bigl( \lambda^{2k}A(\lambda)B(\mu)+\mu^{2k}A(-\mu)B(\lambda)\Bigr)
-2\eta\rho_k(\lambda) B(\lambda)B(\mu),
\nn\\
\nn\\
\{A(\lambda),C(\mu)\}_k&=&
-\frac{\scriptstyle \eta}{\scriptstyle\lambda-\mu}
\Bigl( \mu^{2k}A(\lambda)C(\mu)-\lambda^{2k}A(\mu)C(\lambda)\Bigr)-2\eta\,k\,\lambda A(\lambda)C(\mu)\label{poi2}\\
&-&\frac{\scriptstyle \eta}{\scriptstyle\lambda+\mu} \Bigl( \mu^{2k}A(\lambda)C(\mu)+\lambda^{2k}A(-\mu)C(\lambda)\Bigr)+2\eta\rho_k(\lambda) B(\lambda)C(\mu),
\nn\\
\nn\\
\{B(\lambda),C(\mu)\}_k&=&\frac{\scriptstyle \eta\lambda^{2k}}{\scriptstyle \lambda-\mu}\Bigl(A(-\lambda)A(\mu)-A(\lambda)A(-\mu)\Bigr)\nn\\
&+&\frac{\scriptstyle \eta\lambda^{2k}}{\scriptstyle \lambda+\mu}\Bigl(A(\lambda)A(\mu)-A(-\lambda)A(-\mu)\Bigr)\nn\\
&+&2\eta\Bigl(\rho_k(\mu)A(-\mu)-\rho_k(-\mu)A(\mu)\Bigr)B(\lambda).\nn
\en}
Here $k=0,1$ and $\rho_k=\left[ \frac{\lambda^{2k-1} A(\lambda)}{B(\lambda)} \right]$ is
the quotient of polynomials in  variable $\lambda$ over a field, such that
\[
\rho_0=0,\qquad\mbox{and}\qquad \rho_1=\alpha\lambda+A_{2n}.
\]
\end{prop}
\par\noindent
\textsc{Proof:}
It is sufficient to check the statement on an open dense subset of the reflection equation algebra defined by the assumption that $A(\lambda)$ and $B(\lambda)$ are co-prime and all double roots of  $B(\lambda)$ are distinct.

This assumption allows us to construct a separation representation for the reflection equation  algebra (\ref{rrpoi}). In this special representation one has $n$ pairs of Darboux variables, $\lambda_i$, $\mu_i$, $i=1,\ldots,n$, having the standard Poisson brackets,
\bq
\left\{\lambda_i,\lambda_j\right\}_0=\left\{\mu_i,\mu_j\right\}_0=0,\qquad \left\{\lambda_i,\mu_j\right\}_0=\delta_{ij},
\label{Darb}
\eq
with the $\lambda$-variables being $n$ zeros of the polynomial $B(\lambda)$ and the $\mu$-variables being values of the polynomial $A(\lambda)$ at those zeros,
\bq
B(\pm\lambda_i)=0,\qquad \mu_i=\eta^{-1}\ln A(\lambda_i),\qquad i=1,\ldots,n.
\label{dn-var}
\eq
The interpolation data (\ref{dn-var}) plus $n+2$ identities
\[A(\lambda_i)A(-\lambda_i)=-d(\lambda_i),\qquad A(0)=\sqrt{Q_0},\qquad A= \alpha\lambda^{2n+1}+\ldots,\]
allow us to construct the needed separation representation for the whole algebra:
\ben
\label{DN-rep}
B(\lambda)&=&\displaystyle \lambda\prod_{k=1}^n(\lambda^2-\lambda_k^2),\\
\nn\\
A(\lambda)&=&
\displaystyle
\left( \alpha\lambda+\frac{\scriptstyle (-1)^n\sqrt{Q_0}}{\scriptstyle  \prod \lambda_k^2 }
\right)\prod_{k=1}^n (\lambda^2-\lambda_k^2)
+\sum_{k=1}^n
\left[
 \dfrac{\lambda(\lambda-\lambda_k)\e^{\eta\mu_k}}{2\lambda_k^2}
+\dfrac{\lambda(\lambda+\lambda_k)\e^{-\eta\mu_k}}{2\lambda_k^2}
\right]
\nn\\
\nn\\
C(\lambda)&=&\dfrac{A(\lambda)A(-\lambda)-d(\lambda)}{B(\lambda)}\,.\nn
\en
Using this representation we can easy calculate the bracket $\{.,.\}_1$ (\ref{poi2}) in ($\lambda,\mu$)-variables
\bq
\left\{\lambda_i,\lambda_j\right\}_1=\left\{\mu_i,\mu_j\right\}_1=0,\qquad \left\{\lambda_i,\mu_j\right\}_1=\lambda_i^2\delta_{ij},
\label{NDarb}
\eq
In order to complete the proof we have to check that brackets (\ref{NDarb}) is compatible with the canonical brackets (\ref{Darb}).
The compatibility of the brackets (\ref{Darb}),(\ref{NDarb}) implies the compatibility of the brackets (\ref{rrpoi}),(\ref{poi2}) and vice versa. This completes the proof.

\begin{rem}
The coefficients of the determinant
$d(\lambda)$ (\ref{Acentre})
are the Casimir functions for the both brackets $\{.,.\}_0$ and $\{.,.\}_1$. It means that
the Poisson bracket $\{.,.\}_1$ has the same foliation by symplectic leaves as $\{.,.\}_0$.
\end{rem}

\begin{prop}
The brackets (\ref{poi2}) may be rewritten in the following $r$-matrix form
\ben
\{\,\on{T}{1}(\lambda),\,\on{T}{2}(\mu)\}_k&=& r_{12}^{(k)}(\lambda,\mu)\,
\on{T}{1}(\lambda)\on{T}{2}(\mu)- \on{T}{1}(\lambda)\on{T}{2}(\mu)\,r_{21}^{(k)}(\lambda,\mu)
\label{rrpoi2}\\
&+&
\on{T}{1}(\lambda)\,s_{12}^{(k)}(\lambda,\mu)\,\on{T}{2}(\mu)-\on{T}{2}(\mu)\,s_{21}^{(k)}(\lambda,\mu)\,
\on{T}{1}(\lambda)
\,, \nn
\en
where
\ben
r_{12}^{(k)}(\lambda,\mu)&=&-\eta
\left(\begin{smallmatrix}
 \frac{\lambda^{2k+1}+\mu^{2k+1}}{\lambda^2-\mu^2} & 0 & 0 & 0 \\
 0& 0 & \frac{\mu^{2k}}{\lambda-\mu} & 0 \\
 0 & \frac{\lambda^{2k}}{\lambda-\mu}  & 0 & 0 \\
 0 & 2\rho_k(-\mu) &  -2\rho_k(-\lambda) & \frac{\lambda^{2k+1}+\mu^{2k+1}}{\lambda^2-\mu^2}
\end{smallmatrix}\right)\nn\\
\nn\\
r_{21}^{(k)}(\lambda,\mu)&=&-\eta
\left(\begin{smallmatrix}
 \frac{\lambda^{2k+1}+\mu^{2k+1}}{\lambda^2-\mu^2} & 0 & 0 & 0 \\
  -2\rho_k(\lambda)& 0 & \frac{\mu^{2k}}{\lambda-\mu} & 0 \\
  2\rho_k(\mu) & \frac{\lambda^{2k}}{\lambda-\mu}  & 0 & 0 \\
 0 & 0 & 0 & \frac{\lambda^{2k+1}+\mu^{2k+1}}{\lambda^2-\mu^2}
\end{smallmatrix}\right)\label{rk}\\
\nn\\
s_{12}^{(k)}(\lambda,\mu)&=&-\eta
\left(\begin{smallmatrix}
 \frac{\lambda^{2k+1}-\mu^{2k+1}}{\lambda^2-\mu^2} & 0 & 0 & 0 \\
  -2\rho_k(\lambda) & 0 & \frac{\mu^{2k}}{\lambda+\mu} & 0 \\
 0 & \frac{\lambda^{2k}}{\lambda+\mu}  & 0 & 0 \\
 0 &-2\rho_k(-\mu) & 0& \frac{\lambda^{2k+1}-\mu^{2k+1}}{\lambda^2-\mu^2}
\end{smallmatrix}\right)\nn
\en
and
\[
s_{21}^{(k)}(\lambda,\mu)=\Pi s_{12}^{(k)}(\mu,\lambda) \Pi.\nn
\]
\end{prop}
The proof consists of the straightforward calculations.

Using the separated representation (\ref{DN-rep}) we can rewrite the higher order Poisson brackets
\bq
\left\{\lambda_i,\lambda_j\right\}_k=\left\{\mu_i,\mu_j\right\}_k=0,\qquad \left\{\lambda_i,\mu_j\right\}_k=\lambda_i^{2k}\delta_{ij},\qquad k=0,\ldots,n
\eq
at the $r$-matrix form (\ref{rrpoi2}). As a result we obtain a family of the Poisson brackets compatible with the bracket (\ref{rrpoi}). For the Sklyanin $r$-matrix algebra such family of the brackets
has been constructed in \cite{ts07}.

\begin{prop}  Integrals of motion $H_i$ from {\rm tr}$K(\lambda)T(\lambda)$ are in the  bi-involution
\[\left\{H_i,H_j\right\}_0=\left\{H_i,H_j\right\}_1=0,\]
with respect to the brackets (\ref{poi2}) or (\ref{rrpoi2}).
\end{prop}
\textsc{Proof:} According to \cite{sokts}   variables $\lambda_i,\mu_i$ (\ref{DN-rep}) are the
separated coordinates for the coefficients $H_i$ of the polynomial $\tau$=tr$K(\lambda)T(\lambda)$ and  the separated relations look like
\[
\mbox{\rm tr}K(\lambda_i)T(\lambda_i)=\sum \tau_k\lambda_i^k=\mathcal A(\lambda_i)e^{\eta\mu_i}+\mathcal D(\lambda_i)e^{-\eta\mu_i},\qquad i=1,\ldots,n.
\]
Remind that $\mathcal A(\lambda), \mathcal D(\lambda)$ are numerical polynomials and
among all the coefficients $\tau_k$ we have only $n$ integrals of motion $H_i$.

On the other hand from (\ref{NDarb}) follows that $\lambda_i,\mu_i$ are the Darboux-Nijenhuis variables for the brackets $\{.,.\}_{0,1}$ (\ref{poi2}). So,
integrals of motion $H_i$ are in the bi-involution with respect to the brackets  (\ref{poi2}) or (\ref{rrpoi2}) according to the Theorem 3.2 from \cite{fp02}. This completes the proof.

Summing up, we have proved a bi-involution of the  integrals of motion $H_i$ using the Darboux-Nijenhuis variables $\lambda_i,\mu_i$ and the separation representation  (\ref{DN-rep}) for the reflection equation algebra.

\section{The Heisenberg magnet}
\setcounter{equation}{0}
Another important representations of the quadratic Poisson algebra with the generators $A_{i},B_{i},$ $C_{i}$ comes as a consequence of the
co-multiplication property of the reflection equation algebra (\ref{rrpoi}). Essentially, it means that the matrix $T(\lambda)$ (\ref{22T}) can be factorized into a product of elementary matrices, each containing only one degree of freedom \cite{skl88}. In this picture, our main model turns out to be an $n$-site Heisenberg magnet with boundary conditions, which is an integrable lattice of $n$ sl(2) spins with nearest neighbor interaction.

According to  \cite{skl88} the $2\times 2$ Lax matrix for the generalized Heisenberg magnet acquires the form
\bq\label{22mag-ref} \mathcal T=K_+(\lambda)T(\lambda),
\qquad\mbox{\rm where}\quad
K_+=\left(
 \begin{array}{cc}
 b_1\lambda+b_0\quad & 0 \\
 \lambda\quad & -b_1\lambda+b_0
 \end{array}
 \right),
\eq
and matrix
\bq\label{mag22}
T(\lambda)=\left(\prod_{m=1}^n L_m(\lambda)\right)K_-(\lambda)
\left( \prod_{m=1}^n L_m(-\lambda)\right)^{-1}
 \,,
\eq
with
\[
L_m(\lambda)=\left(\begin{matrix}
\lambda-s_3^{(m)}& s_1^{(m)}+\mathrm{i}s^{(m)}_2\\
s_1^{(m)}-\mathrm{i}s^{(m)}_2&\lambda+s_3^{(m)}
\end{matrix}\right),\qquad
K_-(\lambda)=\left(
 \begin{array}{cc}
 a_1\lambda+a_0 & \lambda \\
 0 & -a_1\lambda+a_0
 \end{array}
 \right),\]
satisfies to the reflection equation algebra at $\alpha=a_1$ and $\eta=-\mathrm i$.
Here $a_{0,1}$, $b_{0,1}$ and $c_m$ are arbitrary numbers, $\mathrm i^2=-1$ and $s_3^{(m)}$ are dynamical variables on the direct sum of sl(2)
\bq \label{P-sl2}
\left\{s_i^{(m)},s_j^{(m)}\right\}_0=\varepsilon_{ijk}\,s_k^{(m)}\,,
\eq
where $\varepsilon_{ijk}$ is the totally skew-symmetric tensor.

Substituting matrix $T(\lambda)$ (\ref{mag22}) into the brackets $\{.,.\}_k$ (\ref{poi2})  one get the overdetermined system of algebraic equations on the elements of the Poisson bivector $P_1$ associated with the Poisson brackets $\{.,.\}_1$:
\bq \label{pois-br}
 \{f(z),g(z)\}_1=\langle df,\,P_1\,dg \rangle=\sum_{i,k}
P_1^{ik}(z)\dfrac{\partial f(z)}{\partial z_i}\dfrac{\partial
g(z)}{\partial z_k}\,,
\eq
where $z=(z_i,\ldots,z_m)$ are coordinates on the Poisson manifold. In our case $z$ consist of coordinates $s_i^{(m)}$ on $n$ copies of sl(2) (\ref{P-sl2}). Solving this system of equations we obtain the second cubic bracket $\{.,.\}_1$ compatible with (\ref{P-sl2}). As an example the local brackets look like
\[
\left\{s_i^{(m)},s_j^{(m)}\right\}_1=\varepsilon_{ijk}\,s_k^{(m)}\Bigl(
(s_3^{(m)}+c_m)^2-2a_1(s_1^{(m)}+\mathrm is_2^{(m)})(s_3^{(m)}+c_m)-2a_0(s_1^{(m)}+\mathrm is_2^{(m)})\Bigr).
\]
For the sake of brevity we omit the explicit form of the nonlocal cubic brackets $\left\{s_i^{(m)},s_j^{(\ell)}\right\}_1$ in this paper.

\section{The generalized Toda lattices}
\setcounter{equation}{0}
The Toda lattices appear as another specialization of our basic model.
Let us consider generalized open Toda lattices with the Hamiltonians
\bq\label{ham-BCD}
H_g=\sum_{i=1}^n {p_i}^2+2\sum_{i=1}^{n-1}
e^{q_i-q_{i+1}}+V_g(q),\qquad \mbox{\rm where} \qquad
\left\{\begin{array}{l}
 V_{\mathscr B}=2a_0 \e^{q_n} \\ \\
 V_{\mathscr C}=a_1^2 \e^{2q_n}\\ \\
 V_{\mathscr D}=2a_2^2\e^{q_{n-1}+q_n}
 \end{array}.\right.
\eq
These Toda lattices are associated with the root systems $\mathscr B_n$, $\mathscr C_n$ and $\mathscr D_n$ \cite{bog76}.

According to  \cite{skl88,sokts} the $2\times 2$ Lax matrix for the generalized open Toda lattice acquires the form
\bq\label{22toda-ref} \mathcal T_{open}=K_+(\lambda)T(\lambda),
\qquad\mbox{\rm where}\quad
K_+=\left(
 \begin{array}{cc}
 0\quad & 0 \\
 1\quad & 0
 \end{array}
 \right),
\eq
and matrix
\bq\label{toda22}
T(\lambda)=\left(\prod_{k=1}^n L_k(\lambda)\right)K_-(\lambda)
\left( \prod_{k=1}^n L_k(-\lambda)\right)^{-1}
 \,,
\eq
with
\[L_i=\left(\begin{array}{cc}\lambda-p_i &\, -\e^{q_i} \\
 \e^{-q_i}& 0
\end{array}\right),\qquad
K_-(\lambda)=\left(
 \begin{array}{cc}
 2a_2\lambda^2-ia_1\lambda+a_0 & (4a_2\e^{q_n}+1)\lambda \\
 0 & 2a_2\lambda^2+ia_1\lambda+a_0
 \end{array}
 \right),\]
satisfies to the reflection equation algebra at $\alpha=0$ and $\eta=1$.
Here $p_i,q_i$ are dynamical variables and $a_k$ are parameters.

The polynomial
\[\mbox{\rm tr}\,\mathcal T_{open}=\lambda^{2n+1}+\sum_{i=1}^n H_i\lambda^{2(n-i)}\]
is a generating function of independent integrals of motion $H_i$.
The first integral $H_1$ coincides with one of the the Hamiltonians (\ref{ham-BCD}), but for the Toda lattices of $\mathscr C_n$ and $\mathscr D_n$ type
we have to change variables
\bq\label{can1}
p_n\to p_n-ia_1e^{q_n}.
\eq
and
\bq\label{can2}
p_n\to -p_n\frac{\cosh(q_n)+1}{\sinh(q_n)},\qquad
q_n\to-\ln(-2a_2\cosh\bigl(q_n+\ln(-a_2)\bigr)+1)\,.
\eq
respectively.

Substituting matrix $T(\lambda)$ (\ref{toda22}) into the brackets $\{.,.\}_k$ (\ref{poi2})  one can rewrite the Poisson brackets $\{.,.\}_k$ in $(p,q)$ variables. Of course, at $k=0$ we obtain canonical bracket
\[
\{q_i,q_j\}_0=\{p_i,p_j\}_0=0,\qquad \{q_i,p_j\}_0=\delta_{ij}.
\]
For the Toda lattice associated with $\mathscr{BC}_n$ root system
after canonical transformation (\ref{can1})  one get the following non-zero brackets
\bq\label{BC-P}
\begin{array}{ll}
i<j\qquad&
\{q_i,q_j\}_1=2p_i,\\
\\
i=1,\ldots, n-1\qquad
&\{p_i,p_{i+1}\}_1=-(p_i + p_{i+1})\,e^{q_i -q_{i+1}},\quad\{q_{i+1},p_i\}_1=e^{q_i-q_{i+1}},\\
\\
&\{q_i,p_i\}_1=p_i^2+2e^{q_i-q_{i+1}},\\
\\
i=1,\ldots,n-2& \{q_i,p_{i+1}\}_1=2e^{q_{i+1}-q_{i+2}}-e^{q_i-q_{i+1}},\\
\\
&\{p_n,q_i\}_1=2e^{q_{n-1}-q_n}-2a_0e^{q_n}-2a_1^2e^{2q_n},\\
\\
i=3,\ldots,
n-1&\{p_i,q_j\}_1=2e^{q_{i-1}-q_i}-2e^{q_i-q_{i+1}},\qquad 1\leq
j\leq i-2,\\\end{array} \eq
and
\[
\{q_n,p_n\}_1=p_n^2+2a_0e^{q_n}+a_1^2e^{2q_n},\qquad
\{q_{n-1},p_n\}_1=2a_0e^{q_n}+2a_1^2e^{2q_n}-e^{q_{n-1}-q_n}\,.
\]
For the Toda lattice associated with $\mathscr{D}_n$ root system after canonical transformation (\ref{can2}) one gets the following non-zero brackets
\bq\label{D-P}
\begin{array}{ll}
i<j\qquad&
\{q_i,q_j\}_1=2p_i,\\
\\
i=1,\ldots, n-2\qquad
&\{p_i,p_{i+1}\}_1=-(p_i + p_{i+1})\,e^{q_i -q_{i+1}},\quad\{q_{i+1},p_i\}_1=e^{q_i-q_{i+1}},\\

\\
&\{q_i,p_i\}_1=p_i^2+2e^{q_i-q_{i+1}},\\
\\
&\{p_n,q_{i}\}_1=2e^{q_{n-1}-q_n}-2a_2^2e^{q_{n-1}+q_n},\\
\\
i=1,\ldots,n-3& \{p_{n-1},q_i\}_1=2e^{q_{n-2}-q_{n-1}}-2e^{q_{n-1}-q_{n}} -2a_2^2e^{q_{n-1}+q_{n}},\\
\\
i=3,\ldots, n-1&\{p_i,q_j\}_1=2e^{q_{i-1}-q_i}-2e^{q_i-q_{i+1}},\quad
1\leq j\leq i-2,
\end{array}
\eq
and
\[\begin{array}{l}
\{q_{n},p_n\}_1=p_n^2\,,\qquad\{q_{n-2},p_{n-1}\}_1=-2e^{q_{n-2}-q_{n-1}}
+2e^{q_{n-1}-q_n}+2a_2^2e^{q_{n-1}+q_n}\,,\\
\\
\{q_{n-1},p_{n-1}\}_1=p_{n-1}^2+2e^{q_{n-1}-q_n}+2a_2^2e^{q_{n-1}+q_n},\\
\\ \{q_n,p_{n-1}\}_1=\{p_n,q_{n-1}\}_1=e^{q_{n-1}-q_n}-a_2^2e^{q_{n-1}+q_n},\\
\\
\{p_{n-1},p_n\}_1=-(p_{n-1} + p_{n})\,e^{q_{n-1} -q_{n}}+(p_{n+1} -
p_{n})\,a_2^2e^{q_{n-1}+q_{n}}\,.
\end{array}
\]
So, using bracket (\ref{poi2}) at $k=1$ we  recovered all the known second brackets for the ${\mathscr BC}_n$ and $\mathscr D_n$ Toda lattices \cite{dam04}.

\begin{rem}
According to \cite{kuz97,sokts} if $a_k\neq 0$ then after more complicated
canonical transformation of $p_n$ and $q_n$  we can describe generalized Toda lattice with the following potential
\[
V_g=2a_2^2e^{q_{n-1}+q_n}+\frac{a_1}{\sinh^2q_n}+\frac{2a_0}{\sinh^2(q_n/2)}\,,
\]
which was discovered by Inozemtsev  \cite{in89}. Of course,
the bracket (\ref{poi2}) gives rise the second Poisson structure for this system.
\end{rem}
\begin{rem}
According to \cite{kuzts89a,kuz97,skl88,sokts} if we multiply matrix $T(\lambda)$ (\ref{toda22})
on the matrix
\[K_+(\lambda)=\left(
 \begin{array}{cc}
 2b_2\lambda^2-ib_1\lambda+b_0 & 0 \\
 (4b_2\e^{-q_1}+1)\lambda & 2b_2\lambda^2+ib_1\lambda+b_0
 \end{array}
 \right)\]
the $\tau(\lambda)=$tr$K_+(\lambda)T(\lambda)$ are generating function of integrals of motion for the periodic Toda lattices associated with all the classical root systems. The Darboux-Nijenhuis variables $\lambda_i,\mu_i$ (\ref{DN-rep}) are the separated coordinates for these integrals and, therefore, integrals of motion for the periodic Toda lattices are in bi-involution with respect to the same brackets (\ref{poi2}) or (\ref{BC-P}-\ref{D-P}). All the details may be found in \cite{ts06b}.
\end{rem}

\section{The Kowalevski top on $so^*(4)$}
\setcounter{equation}{0}
Let us consider the Kowalevski top on $so^*(4)$ with the Hamilton function
\begin{equation}\label{h}
H_1=J_1^2+J_2^2+2J_3^2-2bx_1,
\end{equation}
and the second integral of motion
\begin{eqnarray}
H_2=(J_+^2+2bx_+-{\mathcal P}b^2)(J_-^2+2bx_--{\mathcal P}b^2),\label{k}\\
J_\pm=J_1\pm \mathrm iJ_2,\qquad x_\pm=x_1\pm \mathrm ix_2, \qquad (\mathrm i^2=-1),
\nonumber
\end{eqnarray}
which are Poisson commuting
\begin{equation}\label{0}
\{H_1,H_2\}_0=0
\end{equation}
with respect to the following Poisson brackets
\begin{equation}\label{pb-o4}
\{J_i,J_k\}_0=\varepsilon_{ikl} J_l, \qquad\{J_i,x_k\}_0=\varepsilon_{ikl} x_l,\qquad
\{x_i,x_k\}_0=-{\cal P}\varepsilon_{ikl} J_l.
\end{equation}
Here  $\varepsilon_{ikl}$ is the completely antisymmetric tensor, ${\mathcal P}$ is a complex (or real) parameter,  $x=(x_1,x_2,x_3)$ and $J=(J_1,J_2,J_3)$ are coordinates on the Poisson manifold $so^*(4)$.
The Casimirs of the bracket (\ref{pb-o4}) have the form
\begin{equation}\label{caz-o4}
\mathcal C_1=x_1J_1+x_2J_2+x_3J_3,\qquad \mathcal C_2=x_1^2+x_2^2+x_3^2-{\cal P}(J_1^2+J_2^2+J_3^2).
\end{equation}

According to  \cite{kuz02} the $2\times 2$ Lax matrix for the Kowalevski top acquires the form
\bq\label{kow-ref}
\mathcal T=K_+(\lambda)\,T(\lambda),
\qquad\mbox{\rm where}\quad
K_+=\left(
 \begin{array}{cc}
 1\quad & 0 \\
 \lambda\quad & 1
 \end{array}
 \right),
\eq
and matrix $T(\lambda)$ is the following representation of the reflection equation algebra at $n=2$, $\alpha=0$ and $\eta=2\mathrm i$:
{\setlength\arraycolsep{1pt}
\begin{eqnarray}\label{ca66}
A_4&=&-\frac12\left( X^2+J_3^2-{\mathcal P}b^2/2\right),\quad
A_3=-\frac{\mathrm i}{2}\left( X^3 +(J_3^2+bx_1-{\mathcal P}b^2)X+bx_2J_3 \right),\nn\\
A_2&=&\frac{J^2}{2}\left( X^2+J_3^2 \right)+bx_2J_3X-bx_1J_3^2+\frac{b^2}{2}
\left(  {\mathcal P}J_3^2-x_3^2+\frac{\mathcal C_2}{2}\right),\quad
A_0=-\frac{b^2\mathcal C_1^2}{4},\nn\\
A_1&=&\frac{\mathrm i}{2}\left(  J^2X^3+2bx_2J_3X^2
+\left( J_3^2(J^2-bx_1)+b^2(x_2^2-x_3^2
-{\mathcal P}J_1^2) \right.\right.\nn\\
&&\left.\left. +bJ_1(x_1J_1+x_2J_2)\right)X+bx_2J_3(J_2^2+J_3^2-bx_1)+J_1J_2J_3
(bx_1-{\mathcal P}b^2)\right),\nonumber\nn\\
\nn\\
B_3&=&X^2-J^2+2bx_1-{\mathcal P}b^2,\label{ca66}\\
B_1&=&-J^2X^2-2b(x_2J_3-x_3J_2)X-2b\mathcal C_1 J_1-b^2(x_2^2+x_3^2-{\mathcal P}J_1^2),\nn\\
\nn\\
C_3&=&\frac14 \left(X^2+J_3^2\right)\left(X^2+J_3^2-{\mathcal P}b^2\right),\nn\\
C_1&=&-\frac14\left(  J^2X^4+2bx_2J_3X^3
+\left(2J_3^2(J^2-bx_1+{\mathcal P}b^2/2)+\mathcal C_2b^2-b^2(2x_3^2+x_1^2)
\right)X^2\qquad\right.\nn\\
&&\left. +2bx_2J_3(J_3^2-bx_1)X+J_3^4(J^2-2bx_1)
-b^2J_3^2(x_3^2-x_1^2+{\mathcal P}(J_1^2+J_2^2))
+{\mathcal P}b^4x_3^2\right),\nn
\end{eqnarray}
}
where
\begin{equation}\label{ca67}
X=\frac{J_1J_3+bx_3}{J_2},\qquad J^2=J_1^2+J_2^2+J_3^2.
\end{equation}
In contrast with  \cite{kuz02} we use the transposed matrices $K$ and $T$ in (\ref{kow-ref}).

Substituting this representation of the reflection equation algebra into the brackets $\{.,.\}_1$ (\ref{poi2}) and solving the resulting system  of algebraic equations  one gets the following Poisson brackets $\{.,.\}_1$ in $(x,J)$ variables:
{\setlength\arraycolsep{1pt}
\ben
\{J_1,J_2\}_1&=& -J_3X^2-b(x_3J_1+x_2X),
\nn\\
\{J_1,J_3\}_1&=&J_2X^2-\frac{b(x_3J_2-x_2J_3)X}{J_2}+\frac{bx_2(2J_1J_2+bx_2)}{J_2},
\nn\\
\{J_2,J_3\}_1&=&-J_1X^2-b(x_1J_1+x_3J_3)-\frac{bx_2(J_1^2-J_3^2+bx_1)}{J_2},
\nn\\
\{x_1,J_1\}_1&=&-\frac{2Q_1\bigl(b(x_2J_2+x_3J_3)+(J_2^2+J_3^2)J_1\bigr)}{J_2^2}
,
\nn\\
\{x_1,J_2\}_1&=&-x_3X^2+b\mathcal PJ_2X+\frac{Q_1(2J_1^2-H_1)-bx_2x_3J_1}{J_2},
\nn\\
\{x_1,J_3\}_1&=&x_2X^2-b\mathcal P(J_1J_2+bx_2)+\frac{Q_1( J_1X+J_2J_3)+bx_2^2J_1}{J_2},
\nn\\
\nn\\
\{x_2,J_1\}_1&=&-x_3J_1^2-\frac{(bx_2+J_3X+J_1J_2)\bigl((b\mathcal P-2x_1)J_3+x_3J_1\bigr)}{J_2},
\nn\\
\{x_2,J_2\}_1&=&-\frac{Q_2(H_1-J_1^2+b^2\mathcal P)+bJ_3(\mathcal PJ_3^2+x_3^2)}{J_2}\label{poi2-kow}
\\
\{x_2,J_3\}_1&=&b\mathcal P(J_1^2+J_3^2)
 +\frac{Q_2(J_2J_3+J_1X) +(b\mathcal P-x_1)(bx_2+J_3X)J_1+bx_2x_3J_3      }{J_2}
\nn\\
\nn\\
\{x_3,J_1\}_1&=&x_2J_1^2+
\frac{\bigl(b(x_2J_2+x_3J_3)+(J_2^2+J_3^2)J_1\bigr)\left(b\mathcal P-2x_1+\frac{x_2J_1}{J_2}\right)}{J_2},
\nn\\
\{x_3,J_2\}_1&=&x_1X^2-b\mathcal P(bx_1+J_1^2-J_3^2)
 +\frac{Q_3(2J_1^2-H_1) +bx_2(x_1J_1+x_3J_3) }{J_2},
\nn\\
\{x_3,J_3\}_1&=&\frac{(Q_3-b\mathcal P J_2)(J_2J_3+J_1X)}{J_2}
-\frac{bx_2(x_1X+x_2J_3)}{J_2}\nn
\en
}
and
\ben
\{x_1,x_2\}_1&=&(Z_1x_3-Z_2J_3)-\frac{b\mathcal PJ_1Q_1}{J_2}\nn\\
\{x_1,x_3\}_1&=&-(Z_1x_2-Z_2J_2)+\frac{b\mathcal PXQ_1}{J_2},
\nn\\
\{x_2,x_3\}_1&=&(Z_1x_1-Z_2J_1)+b\mathcal P(x_1J_1-x_3J_3)-\frac{b\mathcal Px_2(J_1^2-J_3^2)}{J_2}\nn.
\en
Here
\[Q=x\wedge J=[x_2J_3-x_3J_2, x_3J_1-x_1J_3,x_1J_2-J_1x_2]
\] and
\ben
Z_1&=&
x_3J_3-x_2J_2-2x_1J_1+\frac{x_2(J_1^2-2J_3^2)}{J_2}+
\frac{\bigl(J_3(b\mathcal P-2x_1)+x_3J_1\bigr)X}{J_2},
\nn\\
Z_2&=&b^2\mathcal P^2+(H_1-J_3^2)\mathcal P.\nn
\en
Functions $\mathcal C_{1,2}$ (\ref{caz-o4}) are Casimirs with
respect to the both Poisson structures $\{.,.\}_{0,1}$ (\ref{pb-o4}),(\ref{poi2-kow}).
It allows as to obtain the recursion operator $N$ on the symplectic leaves of $so^*(4)$ and the corresponding Darboux-Nijenhuis variables $\lambda_{1,2}$.

On the other hand we can get these Darboux-Nijenhuis variables using the control matrix theory.
Remind, that according to \cite{fp02} the bi-involutivity of integrals of motion
\bq\label{bi-inv}
\{H_1,H_2\}_0=\{H_1,H_2\}_1
=0
\eq
is equivalent to the existence of  the non-degenerate control matrix $F$, such that
\bq
P_1dH_i=P_0\sum_{j=1}^2 F_{ij}\,dH_j,\qquad  i=1,2.
\eq
Here $P_{0,1}$ are the Poisson bivectors associated with the brackets $\{.,.\}_{0,1}$.
In our case $F$ looks like
\[
F=\left(
    \begin{array}{cc}
      \dfrac{H_1+b^2\mathcal P}{2}-X^2-J_3^2 & \dfrac{1}{4} \\ \\
    Z_3& \dfrac{H_1+b^2\mathcal P}{2}
    \end{array}
  \right),
\]
where
\ben
Z_3&=&H_2 -2\bigl(b(b\mathcal P-2x_1)-J_1^2\bigr)X^2+8bx_2J_3X
 +(2\mathcal PJ_3^2-2x_3^2)b^2\nn\\
&+&4bQ_2J_3+2(J_2^2+2J_1^2)J_3^2.\nn
\en
Then we can prove that entry $B(\lambda)$ of the matrix $T(\lambda)$ (\ref{ca66}) coincides with
the characteristic polynomial of  $F$ and, therefore, with the minimal characteristic polynomial
of the recursion operator $N$ on the symplectic leaves of $so^*(4)$.

Summing up, we have proved that roots $\lambda_{1,2}$ of the polynomial $B(\lambda)$ are
the Darboux-Nijenhuis coordinates with respect to  the Poisson structures (\ref{pb-o4})
and (\ref{poi2-kow}). At $\mathcal P=0$ these variables $\lambda_{1,2}$ coincide with the
well-known Kowalevski variables \cite{kuz02}.

\section{The Kowalevski-Goryachev-Chaplygin gyrostat.}
\setcounter{equation}{0}
Let us consider the Kowalevski-Goryachev-Chaplygin gyrostat with
the following Hamilton function
\bq
\begin{array}{c}
 H_1=J_1^2+J_2^2+2J_3^2+2\rho J_3+c_1x_1+c_2x_2+c_3(x_1^2-x_2^2)
 +c_4x_1x_2+\dfrac{\delta}{x_3^2}\\
 \\
c_1,c_2,c_3,c_4,\rho,\delta\in{\mathbb R}\,.
\end{array}
\label{Hkgch}
\eq
Here $x=(x_1,x_2,x_3)$ and $J=(J_1,J_2,J_3)$ are coordinates on the dual Lie algebra
$e^*(3)$ with the Lie-Poisson brackets
\begin{equation} \label{e3}
\bigl\{ J_i\,,J_j\,\bigr\}_0= \varepsilon_{ijk}\,J_k\,, \qquad \bigl\{
J_i\,,x_j\,\bigr\}_0= \varepsilon_{ijk}\,x_k\,,\qquad \bigl\{
x_i\,,x_j\,\bigr\}_0= 0,
\end{equation}
and with the following Casimirs
\begin{equation}\label{caz-e3}
\mathcal C_1=x_1J_1+x_2J_2+x_3J_3,\qquad \mathcal C_2=x_1^2+x_2^2+x_3^2.
\end{equation}
The Hamilton function (\ref{Hkgch}) determines dynamical system on $e^*(3)$,
which is an integrable by Liouville at $\mathcal C_1=0$ only.

Let us start with the $2\times 2$ Lax
matrix for the symmetric Neumann system
\bq
L(\lambda)= \left(
\begin{array}{cc}
 \lambda^2-2J_3\lambda-J_1^2-J_2^2-\dfrac{\delta}{x_3^2}\qquad &
 \lambda(ix_1+x_2)-x_3(iJ_1+J_2) \\
 \\
 \lambda(ix_1-x_2)-x_3(iJ_1-J_2) & x_3^2
\end{array}
\right)\,,\label{Laxneum}
\eq
which is a  representation of the Sklyanin algebra on the subset of $e^*(3)$ defined by $\mathcal C_1=0$ \cite{kuzts89,ts02}. Using a family of the Poisson brackets compatible with the Sklyanin
algebra \cite{ts07} we can get compatible bi-hamiltonian structures on $e^*(3)$ associated with the matrix $L(\lambda)$. All the detail may be found in \cite{kts07,ts07}.

According to \cite{kuzts89,ts02}, the Lax matrix for the
 Kowalevski-Goryachev-Chaplygin gyrostat acquires the form
\[
\mathcal T(\lambda)={K}_+(\lambda)T(\lambda),\]
where
\bq T(\lambda)=\,L(\lambda-\rho)\,{{K}}_-(\lambda)\,\left(\begin{array}{cc}
 0 & 1 \\
 -1 & 0
\end{array}\right)L^T(-\lambda-\rho)\left(\begin{array}{cc}
 0 & 1 \\
 -1 & 0
\end{array}\right),\qquad \rho\in\mathbb R\label{KGC22}
\eq
is the representation of the reflection equation algebra (\ref{rrpoi}) at $n=2$, $\alpha=0$ and
$\eta=2\mathrm i$. Here $L(\lambda)$ is given by (\ref{Laxneum}) and
\bq
 {K}_-=\left(\begin{array}{cc}
 a_1\lambda+a_0 & \lambda \\
 0 & -a_1\lambda+a_0
\end{array}\right),\qquad
{K}_+=\left(\begin{array}{cc}
 b_1\lambda+b_0 & 0 \\
 \lambda & -b_1\lambda+b_0
\end{array}\right)
\label{Kpm}
\eq
are numerical matrices depending on arbitrary parameters $a_{0,1}$ and $b_{0,1}$.

The Hamilton function $\widetilde{H}_{1}$ from
\bq\label{int-KGC}
\mbox{\rm tr}\,\mathcal T(\lambda)=\lambda^6-2\widetilde{H_1}\lambda^4+\widetilde{H_2}\lambda^2
+2a_0b_0(\rho^2\mathcal C_2-\delta)
\eq
coincides with the previous function $H$ (\ref{Hkgch}) after canonical transformation of variables
\[
J\to J+Ux\,,\qquad U=\left(\begin{array}{ccc}
 0 & 0 &\mathrm i\beta_+ \\
 0 & 0 & \beta_-\\
 -\mathrm i\beta_+ & -\beta_-& 0
\end{array}\right)\,,\qquad \beta_\pm=\dfrac{a_1\pm b_1}2
\]
and exchange of parameters
\[a_1^2=\left(c_3+\dfrac{\mathrm ic_4}2\right),\qquad
 b_1^2=\left(c_3-\dfrac{\mathrm ic_4}2\right),\qquad
 a_0=\dfrac{\mathrm ic_1-c_2}2,\qquad b_0=\dfrac{\mathrm ic_1+c_2}2\,.
\]

Substituting matrix $T(\lambda)$ (\ref{KGC22}) into the brackets $\{.,.\}_k$ (\ref{poi2})  one get the second cubic bracket $\{.,.\}_1$ compatible with bracket (\ref{e3}).
For brevity we present these brackets at $c_3=c_4=\rho=\delta=0$ only, i.e. for the Kowalevski top:
\ben
\{J_1,J_2\}_1&=&-(J_3(J_1 - \mathrm i J_2)+ 2a_0x_3)(J_1 +\mathrm iJ_2)\nn\\
\{J_1,J_3\}_1&=&-\bigl(J_1^2+J_2^2+2J_3^2-2a_0(x_1-\mathrm i x_2)\bigr)J_2\nn\\
\{J_2,J_3\}_1&=&\bigl(J_1^2+J_2^2+2J_3^2-2a_0(x_1-\mathrm i x_2)\bigr)J_1\nn\\
\nn\\
\{x_1,J_1\}_1&=&(2\mathrm i J_2^2-2a_0x_2)x_3-2x_2(J_1+\mathrm i J_2)J_3\nn\\
\{x_1,J_2\}_1&=&-\mathrm i (J_1^2-2\mathrm i J_1J_2+J_2^2-2\mathrm i a_0x_2)x_3+2\mathrm i x_2(J_1+\mathrm i J_2)J_3\nn\\
\{x_1,J_3\}_1&=&
-\bigl(J_1^2+J_2^2+4J_3^2-2a_0(x_1+\mathrm i x_2)\bigr)x_2+2x_3J_2J_3
\label{poi2-KGC}\\
\nn\\
\{x_2,J_1\}_1&=&(J_1^2+2\mathrm iJ_1J_2+J_2^2-2a_0x_1)x_3+2x_1(J_1+\mathrm i J_2)J_3,
\nn\\
\{x_2,J_2\}_1&=&2\mathrm i(J_1^2+a_0x_1)x_3-2\mathrm i x_1(J_1+\mathrm i J_2)J_3,
\nn\\
\{x_2,J_3\}_1&=&\bigr(J_1^2+J_2^2+4J_3^2-2a_0(x_1+\mathrm i x_2)\bigl)x_1-2x_3J_1J_3,
\nn\\ \nn\\
\{x_3,J_1\}_1&=&x_2(J_1^2+2\mathrm iJ_1J_2+J_2^2)-2\mathrm ix_1J_2^2,
\nn\\
\{x_3,J_2\}_1&=&-x_1(J_1^2-2\mathrm iJ_1J_2+J_2^2)-2\mathrm ix_2J_1^2,
\nn\\
\{x_3,J_3\}_1&=&2(x_2J_1-x_1J_2)J_3\nn
\en
and
\[
\{x_i,x_j\}_1=2\mathrm i \varepsilon_{ijk} (x_1J_2-x_2J_1-\mathrm i x_3J_3)\,x_k.
\]
At $\mathcal C_1=0$ integrals of motion $\widetilde{H}_{1,2}$ (\ref{int-KGC}) are in the bi-involution with respect to the compatible brackets $\{.,.\}_{0,1}$ and satisfy to the following relations
\bq
P_1d\widetilde{H}_i=P_0\sum_{j=1}^2 F_{ij}\,d\widetilde{H}_j,\qquad  i=1,2,
\eq
where $P_{0,1}$ are the Poisson bivectors associated with the brackets $\{.,.\}_{0,1}$.
and
\[
F=\left(
    \begin{array}{cc}
      2J_1^2+2J_2^2+4J_3^2-2a_0(x_1+\mathrm i x_2) & 1 \\ \\
   2a_0(\mathrm ix_2J_1-\mathrm ix_1J_2+x_3J_3)(J_1+\mathrm i J_2)1-(J_1^2+J_2^2)^2& 0
    \end{array}
  \right).
\]
Of course, entry $B(\lambda)$ of the matrix $T(\lambda)$ (\ref{KGC22}) coincides with
the characteristic polynomial of  $F$ and, therefore, with the minimal characteristic polynomial
of the recursion operator $N$ on the symplectic leaves of $e^*(3)$.

Summing up, we have proved that  at $\mathcal C_1=0$  the Kowalevski top has the polynomial brackets (\ref{poi2-KGC}) in additional to the rational brackets (\ref{poi2-kow})  considered above. These different Poisson structures are related with the different separated variables, which give rise to the different representations of the reflection equation algebra.

\begin{rem} There is another
bi-hamiltonian structure for the Kowalevski top on $e^*(3)$, associated with linear $r$-matrix algebra  \cite{mar98}. According to \cite{ts06} the corresponding Poisson tensor is the rational function $\widehat{P}_1=\mathcal C_1^{-1} P_{pol}$, where $\mathcal C_1$ is the Casimir function (\ref{caz-e3}) and $P_{pol}$ is a cubic polynomial in variables $(x,J)$.

The  Casimir $\mathcal C_2$ does not the Casimir function for the second Poisson bivector $\widehat{P}_1$ from \cite{mar98,ts06}, in contrast with the  second Poisson bivectors associated with the considered above brackets (\ref{poi2-kow}) and (\ref{poi2-KGC}).
\end{rem}

\section{Conclusion}

We present a family of compatible Poisson brackets (\ref{poi2}),(\ref{rrpoi2}), that includes the reflection equation algebra. The application of the $r$-matrix formalism is extremely useful here resulting in drastic reduction of the calculations for a whole set of integrable systems.

For the rational $4\times 4$ matrix $r(\lambda-\mu)$ (\ref{rr}) the similar construction has been applied to other $r$-matrix algebras in \cite{ts07,kts07} and  \cite{ts07b}. It will be interesting to construct the similar families of compatible Poisson brackets associated with the
$4\times 4$ matrices $r(\lambda-\mu)$, which are trigonometric and elliptic functions on spectral parameter.

The research was partially supported by
the RFBR grant 06-01-00140 and grant NSc 5403.2006.1.

\end{document}